\documentclass[Afour,sagev,times]{sagej}

\usepackage{moreverb,url}
\usepackage{subcaption}
\usepackage[table]{xcolor}
\usepackage{multirow} 
\usepackage{stfloats} 
\usepackage{lastpage}
\usepackage[colorlinks,bookmarksopen,bookmarksnumbered,citecolor=red,urlcolor=red]{hyperref}

\newcommand\BibTeX{{\rmfamily B\kern-.05em \textsc{i\kern-.025em b}\kern-.08em
T\kern-.1667em\lower.7ex\hbox{E}\kern-.125emX}}

\def\volumeyear{2016}

\begin{document}

\runninghead{Krug et~al.}

\title{A Highly Configurable Framework for Large-Scale Thermal Building Data Generation to drive Machine Learning Research}

\author{Thomas Krug\affilnum{1}\affilnum{2} 
and Fabian Raisch\affilnum{1}\affilnum{3}
and Dominik Aimer\affilnum{1}
and Markus Wirnsberger\affilnum{1}
and Ferdinand Sigg\affilnum{13}
and Felix Koch\affilnum{1} 
and Benjamin Schäfer\affilnum{2}
and Benjamin Tischler\affilnum{1}}

\affiliation{\affilnum{1}Rosenheim Technical University of Applied Sciences, Germany\\
\affilnum{2}Karlsruhe Institute of Technology, Germany\\
\affilnum{3}Technical University of Munich, Germany}

\corrauth{
Thomas Krug,
Rosenheim Technical University of Applied Sciences,
Hochschulstraße 1,
83024 Rosenheim,
Germany}

\email{thomas.krug@th-rosenheim.de}

\begin{abstract}

Data-driven modeling of building thermal dynamics is emerging as an increasingly important field of research for large-scale intelligent building control. However, research in data-driven modeling using machine learning (ML) techniques requires massive amounts of thermal building data, which is not easily available. Neither empirical public datasets nor existing data generators meet the needs of ML research in terms of data quality and quantity. Moreover, existing data generation approaches typically require expert knowledge in building simulation. To fill this gap, we present a thermal building data generation framework which we call \texttt{BuilDa}. \texttt{BuilDa} is designed to produce synthetic data of adequate quality and quantity for ML research. The framework does not require profound building simulation knowledge to generate large volumes of data. \texttt{BuilDa} uses a single-zone Modelica model that is exported as a Functional Mock-up Unit (FMU) and simulated in Python. We demonstrate \texttt{BuilDa} by generating data and utilizing it for a transfer learning study involving the fine-tuning of 486 data-driven models.

\end{abstract}


\maketitle

\section{Introduction}
\label{sec:intro}

In 2021, building operations accounted for 27\% of global CO$_2$ emissions and 30\% of the world's energy demand \cite{globalstatus}. Intelligent building energy control, such as model predictive control (MPC) \cite{YeYao2021MPCReview, DRGONA2020} or reinforcement learning (RL) \cite{WEINBERG2023_RLCS_Review, Yu2021_RLReview} can significantly reduce building energy consumption. Usually, these methods rely on a representative model of the building's thermal dynamics. Similarly, such models can act as baseline-models for fault detection \& diagnosis (FDD) research \cite{chen2023reviewFDD}, improving energy efficiency. However, building thermal dynamic models often require labor-intensive manual modeling and expert knowledge. Data-driven methods offer a way to overcome this problem, as the models can be learned directly from data \cite{YeYao2021MPCReview, chen2023reviewFDD, ALSAYED2024}. Machine learning (ML) techniques are particularly suitable for this purpose, as they allow to model complex, non-linear building behavior without explicit physical modeling. This makes the wide-spread adoption of intelligent control strategies more feasible. Yet, depending on the task, long-term data collection over several months up to multiple years is often necessary for deriving accurate data-driven models. This is not feasible for most buildings, especially not for recently built ones.

Transfer Learning (TL) as a ML technique addresses the challenge of data scarcity and can improve the robustness of data-driven models \cite{pinto2022transfer, peirelinck2022transfer}. In TL, a model is pretrained with data from a source building to acquire knowledge. Subsequently, the pretrained model is fine-tuned for a target building using a relatively small amount of data measured in the target \cite{chen2020transfer, jiang2019deep, pinto2022transfer}. 
Promising developing research areas in the TL field are research in generalized TL models \cite{2025gentl}, studies on building similarity (e.g., weather conditions, building size, occupancy patterns) \cite{pinto2022sharing, li2024building,2024Similarity, CHAUDHARY2025115384} or studies on the adaptiveness of data-driven models to changes in the built environment\textsuperscript{(anonymous)}. Consequently, advancing research in data-driven building thermal dynamic models and TL is essential in order to achieve a wide-spread adoption of intelligent building applications. This underlines the need for massive amounts of high-variance data with detailed and accessible metadata, which can be tailored to the research questions. 

For this purpose, datasets for ML research should ideally exhibit high variability to cover a wide range of building dynamics, representing numerous buildings and their environmental conditions. Research data should cover all seasons over multiple years to allow for representative training and test sets, while also providing detailed metadata descriptions. Yet, data with desired properties is rare (see Section \ref{sec:background} \nameref{sec:background}), restricting ML research for building thermal dynamics significantly. 
Currently, researchers either need to rely on publicly available datasets or they need to generate synthetic data by themselves. Both options have their own limitations, with the latter often requiring extensive knowledge of building simulation. Researchers, especially in the ML domain, without access to such expertise face significant challenges due to the lack of accessible and detailed data sources.   

\section{Background}
\label{sec:background}

ML research in the building domain often relies on publicly available datasets. These datasets, whether real-world measurements or synthetically generated \cite{miller2020building, miller2017building, li2021synthetic, idealdataset, luo2022ecobee}, typically provide static information that cannot be adapted to different conditions. Since the data is already measured or generated, it represents only a fixed set of buildings and operating scenarios. This lack of flexibility limits detailed analyses of thermal dynamics and constitutes a major drawback of existing datasets.
Moreover, public data sources often lack detailed metadata, especially for occupancy, building envelope characteristics, and weather data. Zhang et al.\cite{2024Similarity} examine how building similarity affects transfer learning performance for cooling load prediction, but couldn't consider features of the building envelope in detail, due to inaccessible metadata. Li et al.\cite{li2024building} use the Ecobee dataset \cite{luo2022ecobee} but had to manually infer weather conditions from the nearest weather station, highlighting data aggregation challenges. Similarly, the IDEAL \cite{idealdataset} dataset faces the same limitation, requiring weather data to be extrapolated from nearby weather stations.  

An alternative approach is the generation of synthetic data using physics-based building models, implemented in simulation environments like Modelica \cite{mattsson1997modelica} or EnergyPlus \cite{2001EnergyPlus}. While these models theoretically enable the generation of unlimited data, they require extensive domain expertise and are time-intensive to set up. Additionally, the models are not often easily transferable across different building types, limiting their adaptability and usefulness for ML research.
Tools such as eppy \cite{eppy} enable systematic modifications of EnergyPlus models but still demand substantial simulation expertise, as parametric adjustments in the model must still be physically consistent.

Another approach involves data generators, which integrate building modeling knowledge, making them more accessible to non-experts. Existing solutions range from data augmentation frameworks \cite{Fan2022SynDataAug} to data generators for synthetic smart meter or electrical data \cite{Hong2020SynSmartMeter, charbonnier2024home}. However, only few focus on thermal building dynamics, such as Synconn\_build \cite{chaudhary2023synconn_build}, SBsim \cite{SBsim} or eplusr \cite{eplusr}. SBsim is limited as it relies on finite-differences approximation of building thermodynamics rather than a full physical simulation model and is primarily intended for RL research. Synconn\_build uses a physical simulation model, but lacks flexibility in terms of adjustable building parameters, limiting its suitability for ML research. Eplusr offers extensive configurability, but still requires expertise in manipulating EnergyPlus IDF files. These challenges underscore the need for accessible, customizable, physically consistent and scalable data-generation frameworks for ML research in building thermal dynamics, particularly for researchers without building simulation expertise.

We present a building data generation framework called \texttt{BuilDa}, designed to generate high-fidelity thermal time-series data for ML research without requiring simulation knowledge. \texttt{BuilDa} is highly customizable, supports parallelized data generation, and utilizes a validated physical building model. Users can adjust weather, system controls, building properties, and operational schedules, including occupancy behavior and retrofitting scenarios. In addition, we provide predefined profiles for climate, wall constructions, internal heat gains and window-opening, and predefined control strategies. This work extends the foundations established in previous research\textsuperscript{ (anonymous)}. 

Our contributions are as follows: (i) We provide an extensive data-generation framework for generating synthetic multivariate thermal building dynamic time series data, which is highly customizable, physically consistent, usable without expertise in building simulation, and suitable for diverse ML research applications. (ii) The framework includes a flexible, validated, high-fidelity single-zone building simulation model for data generation, incorporating extensive building modeling domain knowledge and covering a wide range of possible single-zone buildings. (iii) Moreover, the framework supports the simulation of operational changes and retrofitting scenarios, enabling dynamic modifications of the building envelope during simulation. In addition, we provide predefined load profiles covering occupancy schedules, internal heat gains from occupancy and electrical devices, and correlated window-opening behavior, representing five different household types.
(iv) Finally, we showcase the usage of \texttt{BuilDa} by generating synthetic data for an extended TL study on the influence of building properties on the effectiveness of TL. In the study, 486 standard ML models are pretrained, fine-tuned, and compared.  

The remainder of this paper is structured as follows: Section \ref{cha:Framwork} \nameref{cha:Framwork} introduces the \texttt{BuilDa} framework and its features. Section \ref{cha:demonstration} \nameref{cha:demonstration} presents several demonstration use cases. Thereafter, we conclude the paper in Section \ref{cha:conclusion} \nameref{cha:conclusion}.

\section{Framework and Functionalities}
\label{cha:Framwork}

\subsection{Architecture Overview}
\label{cha:Architecture}

\begin{figure}[!htb]
    \centering
    \includegraphics[width=1\linewidth]{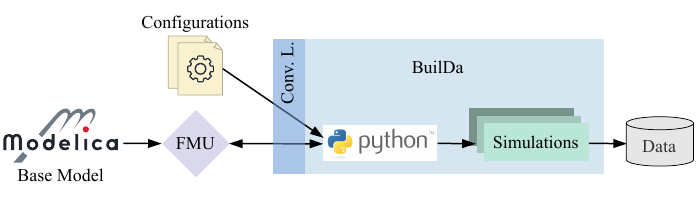}
    \caption{Architecture overview \texttt{BuilDa}.}
    \label{fig:Pipeline}
\end{figure}

This chapter provides an overview of the architecture of the \texttt{BuilDa} framework. As illustrated in \autoref{fig:Pipeline}, the framework consist of two primary components. The first component is a physical Modelica building model (referred to as the base model) that simulates the thermal dynamics of a single-zone building. The second component is a Python-based framework to run multiple simulations with varying parameters and operational schedules. The framework executes the base model as a Functional Mock-up Unit (FMU) via the FMPy library \cite{FMPy}, following the FMI standard 2.0.4 \cite{blochwitz2012functional}. 

Configuration files enable users to set parameters for the building model, control strategy, simulation, and operational schedules. A key feature of \texttt{BuilDa} is the converter layer, which simplifies user input by automatically managing parameter dependencies. The converter layer is described in Subsection \ref{cha:converter_layer} \nameref{cha:converter_layer}. It is specifically tailored to our FMU. Adjustments are required if another FMU with a different parameter set is used. 
Subsection \ref{cha:buildingmodel} \nameref{cha:buildingmodel} provides an overview of base model and the adjustable parameters within \texttt{BuilDa} and Subsection \ref{cha:operational_changes} \nameref{cha:operational_changes} describes how parameters and gain profiles can be dynamically changed during operation.

\subsection{Building Model}
\label{cha:buildingmodel}

\begin{figure}[htb]
    \centering
    \includegraphics[width=1\linewidth]{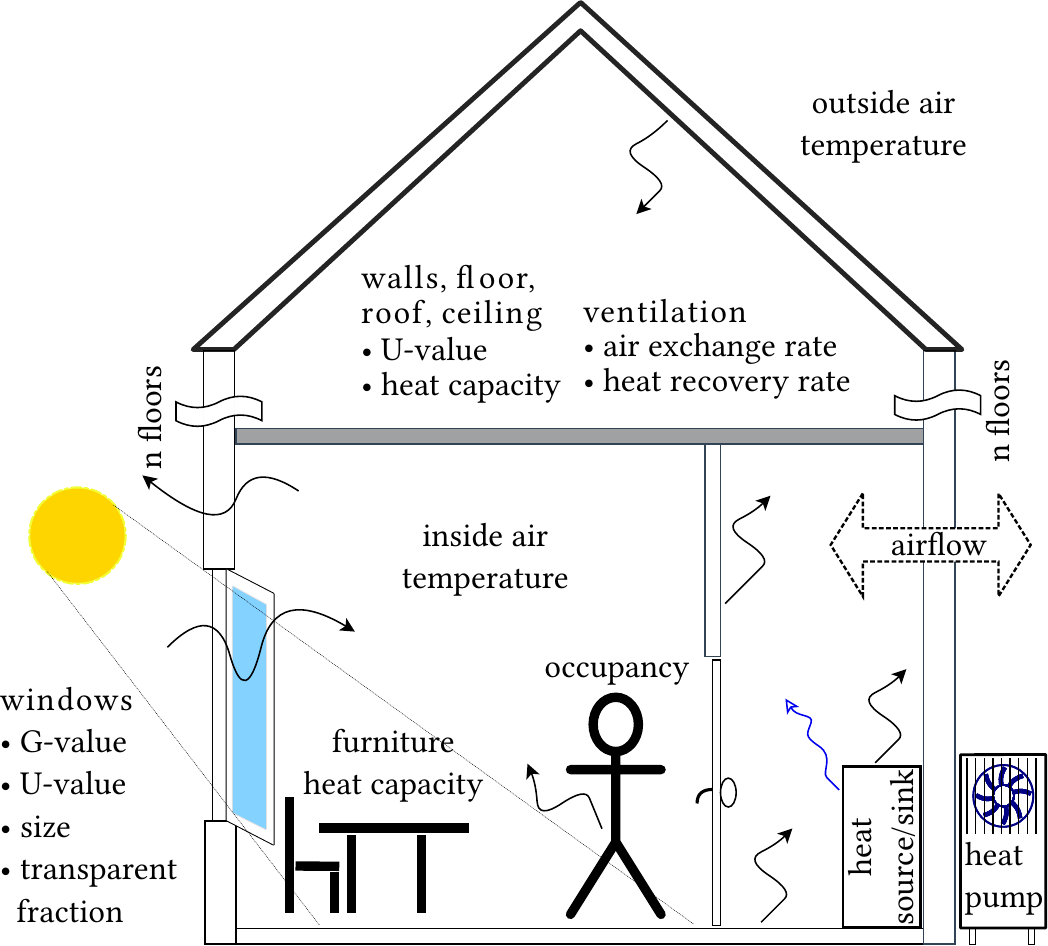}
    \caption{Basic setup of the heating system.}
    \label{fig:setup}
\end{figure}

To ensure model accuracy, the building and its components are modeled in accordance with the methods outlined in VDI 6007 Part 1 for building simulations\cite{VDI6007-1}. \autoref{fig:setup} illustrates the single-zone building model, highlighting its main components and parameters. Heat flows are indicated with arrows (blue for cooling). The model comprises walls, roof, floor, ceiling, windows, and furniture. Heating and cooling are represented by an ideal thermal source with negligible thermal mass, analogous to an electric radiator, which transfers heat directly to the environment. The convective and radiative fractions are defined in the configuration files. Maximum heating power is determined by the converter layer according to DIN 18599 Part 2\cite{DIN18599-2} (see Subsection \ref{cha:converter_layer} \nameref{cha:converter_layer}). The same standard is applied for maximum cooling power, using a simplified approach consistent with the overall model assumptions.

Additionally, a simple heat pump model is included to calculate the electrical power required to supply the current heat flow to the zone. The coefficient of performance (COP) is determined from the characteristic curve of the OVUM ACP520 reference heat pump\cite{OVUM_ACP520_Datenblatt} and depends on the outside temperature, supply temperature, and compressor speed.

Solar gains and thermal losses are considered for windows, exterior walls, and the roof, while only thermal losses are accounted for in the floor. Within the zone, heat exchange between walls and between walls and the thermal source is modeled via long-wave radiation and convection. We further assume a uniform indoor air temperature, which is a valid simplification for many building types, particularly residential buildings. This reduces model complexity while still reflecting typical conditions \cite{TASK44_2013}.

The core model, including thermal zone, envelope, internal gains, and weather, is derived from an example in the Modelica Buildings Library \cite{Wetter014buildinglib}. The model was extended to enable higher configurability and account for aspects such as air exchange, control, heating, cooling, and the heat pump. \autoref{fig:model_overview} in the appendix provides an overview of the Modelica implementation. 

A summary of modifiable parameters is provided in \autoref{tab:building_parameters_short}. Further details are given in the following subsections, where the discussed parameters are highlighted in bold. 

\subsubsection{Building Parameters}
\label{cha:Building Parameters}

These parameters define the characteristics of a building, such as size, envelope quality, thermal mass, windows, and ventilation, all of which affect its thermal behavior. The \textbf{building size} is defined by floor length, width, and height, and the number of floors. Building components—including roof, floor, windows, and interior and exterior walls—are modeled as R–C elements. The R-value describes thermal resistance and thus \textbf{envelope quality}, while the C-value represents \textbf{thermal mass}. Each component is discretized into three R–C nodes, which provide sufficient accuracy for the model \cite{Davies2004}. The achievable accuracy depends on the number of nodes and their distribution within the wall structure. While users can manually configure components by specifying the U-value, heat capacity, and the distribution of resistance and capacity within the component, \texttt{BuilDa} also provides a convenient option to select from predefined component profiles, representing typical configurations. The available profiles are summarized in \autoref{tab:predefined_wall_profiles}.

\begin{table}[!htb]
    \footnotesize
    \centering
    \caption{Predefined wall profiles}
    \label{tab:predefined_wall_profiles}
      \begin{tabular}{p{1.5cm} >{\centering\arraybackslash}p{1.0cm} >{\centering\arraybackslash}p{1.15cm} p{3.15cm}}
        \toprule
        \vspace{0.5em} Name & U-value W/(m²K) & Heat Cap. kJ/(m²K) & \vspace{0.5em} Description \\ 
        \midrule
        \parbox[t]{1.5cm}{High-hole \\ brick (1980s)} & 0.83  & 250  & Old fashioned wall with poor insulation properties, but high inertia \\ 
        Solid brick              & 1.61  & 376  & Old fashioned wall with very poor insulation properties and very high inertia \\ 
        \parbox[t]{1.5cm}{Concrete + \\ETICS}      & 0.21  & 470  & Wall with good insulation properties and high inertia \\ 
        \parbox[t]{1.5cm}{Timber \\ construction}      & 0.15  & 75   & Wall with very good insulation properties and low inertia \\ 
        \parbox[t]{1.5cm}{High-hole \\brick (today)}  & 0.23  & 265  & Wall with good insulation properties and high inertia \\ 
        \parbox[t]{1.5cm}{Drywall}                 & 0.56  & 17.6 & Internal wall with insulation and low heat capacity \\
    \bottomrule
    \end{tabular}   
 \end{table}

\begin{table*}[htb]
\small
\caption{Building simulation parameters.}
\label{tab:building_parameters_short}
\centering
    \begin{tabular}{p{1.3cm}p{2.4cm}p{11.1cm}} 
        \toprule
Category & Parameter & Explanation                                                                                                                             \\ \midrule
Building          & \textbf{Building Size}                               & Thermal envelope size (including length, width, number of floors, and floor height)                                                              \\  
                  & \textbf{Envelope Quality} & Thermal quality of building envelope (U-value, structure)                                                                                        \\  
                  & \textbf{Thermal Mass}                                & Heat capacity of external and internal walls, roof, floor, furniture                                                     \\  
                  & \textbf{Windows}                                     & Size, orientation, and quality of windows (U-value, G-value and transparent fraction)                                                            \\  
                  & \textbf{Ventilation}                                 & Air exchange rates and heat recovery rate                                                                                                        \\ 
                  & \textbf{Heat Pump}                                 & Heat pump efficiency and steepness of heat curve                                                                                                    \\ \hline
Input             & \textbf{Controller}                                  & Configurable external controller(s) (e.g., for heating, cooling, window opening) or internal controller configuration (setpoints)                                          \\ 
                  & \textbf{Internal gains}                                   & Internal gains profile, calculated from occupancy profile (optionally contains heat gains from electrical devices)                                                                                        \\  
                  & \textbf{Window opening}                              & Window opening profile (influences ventilation heat losses), calculated from occupancy profile                                                                                           \\  
                  & \textbf{Weather}                                     & Weather data with outside temperature, solar radiation, etc.\newline (main influence on heat energy demand and zone temperature) \\ 
\bottomrule
\end{tabular}
\end{table*}

The heat capacity of the furniture is considered as part of the internal walls for simplicity, as it's supposed to have a similar impact on the thermal dynamics of the building. For external walls and \textbf{windows}, orientation is taken into account to capture direction-dependent solar gains. The window G-value represents the fraction of solar radiation transmitted through the glazing.

For \textbf{ventilation}, heat flow (due to ventilation system, infiltration, or window opening) is determined by considering the air flows and the heat recovery rate (0 for natural ventilation or exhaust-air systems, $>$0 for ventilation systems with heat recovery). \textbf{Window opening} is included in the model. Following the standard VDI 2078 \cite{VDI2078}, we have modeled the airflow through a fully opened window based on the indoor and outdoor temperatures, as well as the size and dimensions of the window. During the simulation, the window can be opened according to an external window opening profile (see Subsection \ref{cha:input_parameters} \nameref{cha:input_parameters}) or by an external control signal (see \autoref{tab:building_parameters_short}), allowing the user to directly implement customized control logic.

The \textbf{heat pump} in the building model can be configured by adjusting the efficiency in relation to the efficiency of the used reference heat pump \cite{OVUM_ACP520_Datenblatt} (see also Subsection \ref{cha:buildingmodel} \nameref{cha:buildingmodel}). Lastly, the steepness of the heat curve, that maps the outside temperature to the required heating supply temperature can be configured.

\subsubsection{Input Parameters} 
\label{cha:input_parameters}

In contrast to building parameters, input parameters define the dynamic inputs to the model during simulation, including control strategies, occupancy patterns, window-opening behavior, and weather conditions. \texttt{BuilDa} supports two types of \textbf{controllers}. The FMU contains an internal heating controller implemented as a proportional controller with night setback, where day and night setpoints can be configured individually. Alternatively, external controllers can be defined for heating, cooling, and window operation, enabling more flexible and adaptive control strategies. To achieve this level of flexibility, we decoupled the frequency at which the controller updates the actuation signal from the rate at which the simulation retrieves its status and writes data.

Occupancy in \texttt{BuilDa} affects \textbf{internal heat gains} and, optionally, \textbf{window-opening} activity, both represented through annual profiles. A provided Python script can generate these yearly schedules for internal gains (from occupants and electrical devices) and window-opening behavior. The script uses four day profiles (workday, Saturday, Sunday, holiday) with individual activity schedules and optional internal heat gains from electrical devices. Window opening is updated every five minutes based on occupancy, sleep times, and ventilation awareness.

Additionally, \texttt{BuilDa} includes five predefined profile sets, each consisting of an internal gain profile (including heat gains from occupants and electrical devices) and a corresponding window-opening profile, representing typical german residential household archetypes.
Realistic internal heat gains were generated using the \textit{LoadProfileGenerator (LPG)}~\cite{Pfluegl2022LoadProfileGenerator}, which provides detailed stochastic household demand profiles derived from German time-use statistics and appliance models. It simulates person-specific activities and device usage to produce temporally resolved electricity consumption and occupancy data. The selected household archetypes cover different compositions, age groups, and occupancy patterns, ensuring a realistic variety of internal heat gain behaviors throughout the day (see \autoref{tab:predefined_gain_profiles}).

\begin{table}[!htb]
    \footnotesize
    \centering
    \caption{Predefined gain profiles}
    \label{tab:predefined_gain_profiles}
    \begin{tabular}{p{2.6cm}p{5cm}}
        \toprule
        Name & Description \\ 
        \midrule
        \parbox[t]{2.6cm}{\texttt{CHR07\_}\\\texttt{Single\_with\_work}} &
        Single adult, regular daytime work outside home. Typical weekday absence during daytime; 
        morning and evening peaks in appliance use. Represents urban single apartments. \\
        \midrule
        \parbox[t]{2.6cm}{\texttt{CHR01\_}\\\texttt{Couple\_both\_at\_}\\\texttt{Work}} &
        Two adults, both employed. Moderate consumption during weekdays with evening peaks (cooking, laundry, lighting). Represents common dual-income households. \\ 
        \midrule
        \parbox[t]{2.6cm}{\texttt{CHR27\_}\\\texttt{Family\_both\_at\_}\\\texttt{work\_2\_children}} &
        Four-person household (two adults, two school-aged children). Pronounced morning and evening peaks; high weekend variability. Represents family homes in suburban areas. \\ 
        \midrule
        \parbox[t]{2.6cm}{\texttt{CHR16\_}\\\texttt{Couple\_over\_}\\\texttt{65\_years}} &
        Elderly couple, retired. Occupied most of the day; more constant internal heat gains with smaller peak loads. Represents older households or senior residences. \\ 
        \midrule
        \parbox[t]{2.6cm}{\texttt{CHR52\_}\\\texttt{Student\_}\\\texttt{Flatsharing}} &
        Shared flat (3–4 young adults) with irregular schedules and variable appliance use. Captures non-synchronous, high evening loads typical for student housing. \\
    \bottomrule
    \end{tabular}

\end{table}

These profiles offer a diverse representation of occupancy and appliance usage patterns. 
They serve as stochastic input for the thermal model’s internal heat gain calculation, linking electrical power consumption to the corresponding sensible and latent heat releases. Furthermore, heat emissions from occupants are included to account for the metabolic contribution of human presence.

\textbf{Weather} significantly affects a building's thermal behavior. In \texttt{BuilDa}, users can simulate different locations by specifying the weather file path in the configuration file. The model imports weather data in MOS format. An open-source tool converts EPW weather files (available worldwide \cite{TMYx}) to MOS format. \texttt{BuilDa} also includes several weather files for central Europe.

\subsection{Operational Schedules} 
\label{cha:operational_changes} 

During operation, buildings may change due to various factors, such as retrofits of the building envelope or changes in occupancy. This directly affects the thermal dynamic behavior of buildings. \texttt{BuilDa} enables users to specify such modifications by adjusting building envelope parameters or internal gain and window-opening profiles during runtime. For this purpose, a second, optional configuration file is provided. Users can define which parameters or profiles should change at a specific timestamp and whether heating or cooling loads should be recalculated after changes, which is particularly relevant in retrofitting scenarios. \texttt{BuilDa} runs the simulation using the initial configuration until the defined timestamp, applies the specified changes, and then continues until the next change or the end of the run.

\begin{table*}[!htb]
    \small
    \centering
    \caption{Role of different converter functions in \texttt{BuilDa} and their execution order}
    \label{tab:converter_functions}
    \begin{tabular}{clp{9.3cm}}
    \toprule
        Order & Function Name & Explanation \\ \midrule
        1 & \texttt{Link\_Resolver} & 
        Resolving linked parameters: A parameter value can be linked to another so that it automatically adapts its value. For example, the window-to-wall fraction can be defined once for one orientation, while the other orientations are linked to it. This avoids redundant inputs and keeps the configuration simple\\
        2 & \texttt{Miscellaneous\_Handler} & Incorporates functionality witch does not fit in other converter functions. E.g., translates some user parameter names to modelica parameter names. \\
        3 & \texttt{Model\_Compatibility\_Layer} & Prevents conflicts in the model by automatically correcting inconsistent user inputs. For instance, to avoid division-by-zero errors in the FMU, parameters are set to values slightly above zero. \\
        4 & \texttt{Zone\_Dimensions\_Calculator} & Calculates the areas of roof, floor, windows, and exterior and interior walls for each orientation, based on the specified zone length, width, floor height, and number of floors. \\
        5 & \texttt{Component\_Configurator} &  
        Maps the predefined parameter values of wall construction profiles to the respective model parameter values (see \autoref{tab:predefined_wall_profiles}). \\
        6 & \texttt{RC\_Distribution\_Configurator} & Maps predefined parameter values of R and C distribution profiles to the respective model parameter values when this option is enabled. \\
        7 & \texttt{Component\_Properties\_Calculator} & Computes R- and C-value distributions for component RC elements from U-value, heat capacity, element count, and zone dimensions. \\
        8 & \texttt{Nominal\_Heating\_Power\_Calculator} & Calculates the nominal heating power of a building, based on building and environmental parameters (DIN 18599-2). \\
        9 & \texttt{Nominal\_Cooling\_Power\_Calculator} & Calculates the nominal cooling power of a building, based on building and environmental parameters (DIN 18599-2). \\ 
        \bottomrule
    \end{tabular}
\end{table*}

\subsection{Converter Layer}
\label{cha:converter_layer}

The converter layer is a key feature of \texttt{BuilDa} and utilizes the user-defined parameter values to automatically infer missing inputs and resolve parameter dependencies required for simulation. This ensures physical consistency for each individual run. For instance, room area and volume are directly related. In the configuration file, only one value needs to be specified, while the dependent value is computed automatically and mapped to the respective Modelica model parameters. Similarly, cooling and heating power can be automatically recalculated for each building configuration for consistent physical behavior.
The converter functions are described in \autoref{tab:converter_functions}. The functions are executed sequentially for each simulation to handle the correct resolution of the dependencies.

\subsection{Software Usage}
\label{cha:software_usage}

First, the parameter values need to be specified in the configuration file by the user (see appendix \autoref{tab:allParameters} for a list of the major parameters). \texttt{BuilDa} provides various ways to construct a set of variations of the values for simulation, e.g., setting a range of values with a step size or the construction of a cartesian product. Optionally, the operational schedules can be configured as described in Subsection \ref{cha:operational_changes} \nameref{cha:operational_changes}. \texttt{BuilDa} iterates through the set of variations and executes the simulations in parallel, utilizing all available CPU cores. The output is written to the file system as a CSV file, and output columns can be specified during configuration. The duration of a complete simulation run with \texttt{BuilDa} primarily depends on the size of the variation set, the controller's update sampling time, the FMU state output interval, and the number of available compute cores. For instance, simulating the FMU with 100 variations over one year, using an output interval of 900 seconds, required approximately 1240 seconds (averaging 12.4 seconds per simulation) when using the internal controller. In contrast, the same simulation with an external two-point controller, updating the manipulated variable in 900-second steps, took approximately 1600 seconds (averaging 16 seconds per simulation). The simulation was executed on an Intel® Core™ i7-6600U CPU (2.60 GHz, 4 cores) running Ubuntu 22.04 with 20 GB of RAM. The increased execution time in the externally controlled FMU is attributed mainly to additional recalculations within the FMU triggered by each controller update. Lastly, the existing FMU in our framework can be replaced with a different one. As the process is nontrivial, we provide a short description on how to perform the adaptation. The \texttt{BuilDa} framework is available on GitHub\cite{BuilDaRepo}.

\section{Demonstration}
\label{cha:demonstration}

\subsection{Data Showcase}
\label{cha:data_quality}

\begin{table*}[htb]
    \small
    \caption{Example ranges for building parameter values.}
    \centering
    \setlength{\tabcolsep}{4pt} 
    \begin{tabular}{llccl}
\toprule
              Category  & Component       & Min. & Max. & Coverage                                                                                                         \\ \midrule
U-value       & Floor, walls, roof       & 0.1          & 1.4          & directly                                                                                                                         \\
(W/m²/K)      & Window                   & 0.7          & 4.3          & directly                                                                                                                         \\ \midrule
              & Floor                    & 62           & 116          & \textit{zone\_length * zone\_width}                                                                                              \\ 
Area (m²)     & Walls                    & 120          & 230          & \textit{\begin{tabular}[c]{@{}c@{}}2 * n\_floors * floor\_height * (zone\_length + zone\_width) - A\_Windows\end{tabular}} \\  
              & Roof                     & 86           & 185          & \textit{\begin{tabular}[c]{@{}c@{}}zone\_length *  zone\_width * f\_ARoofToAFloor\end{tabular}}                                \\  
              & Window (e.g. south side) & 18           & 42           & \textit{\begin{tabular}[c]{@{}c@{}}floor\_height * n\_floors *  zone\_length * f\_AWin\_south\end{tabular}}                      \\ \hline
Heat capacity & Floor                    & 270          & 500          & directly                                                                                                                         \\
(kJ/m²/K)     & Walls                    & 50           & 660          & directly                                                                                                                         \\
              & Roof                     & 35           & 400          & directly                                                                                                                         \\ \bottomrule
\end{tabular}
    \label{tab:rang_building_para}
\end{table*}

\texttt{BuilDa} can be configured to cover a wide range of parameter values, as shown in \autoref{tab:rang_building_para}. These ranges are provided as guidelines to illustrate possible configurations and are consistent with the representative values of the TABULA database \cite{Loga2011-ws}, which documents German buildings constructed from 1949 to the present. While the parameter ranges are based on German reference data, TABULA databases also exist for other European residential building typologies. It is important to note that the framework is not limited to TABULA values and parameter values can be chosen freely. Most parameters can be specified directly by the user, while others are derived by the converter layer. The framework could in principle also be applied to represent commercial or industrial buildings. However, due to their typically larger size, approximating such buildings with a single-zone model is likely to be of limited utility.

\begin{figure}[!htb]
    \centering
    \subcaptionbox{Simulation of one year. The parameter variation in the legend is listed as follows: external wall's U-value, window-to-wall fraction,
external wall's heat capacity,
location, zone area
    \label{fig:mid_temp_data_diversity}}{
        \includegraphics[width=0.48\textwidth]{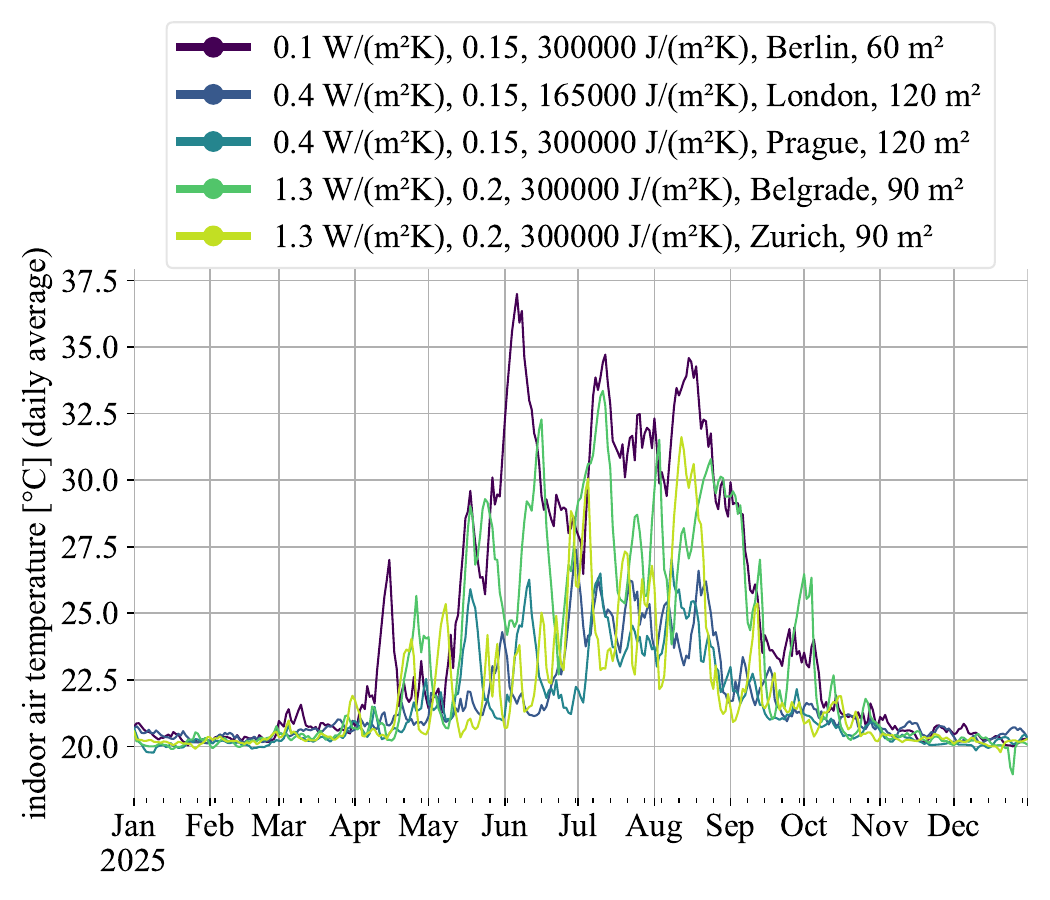}
    }
    \subcaptionbox{Simulation of one day with different controllers. Upper figure: proportional controller. Lower figure: two-point controller. The parameter variation in the legend is listed as follows: external wall's U-value, zone area\label{fig:controller_example_time_series}}{
            \includegraphics[width=0.48\textwidth]{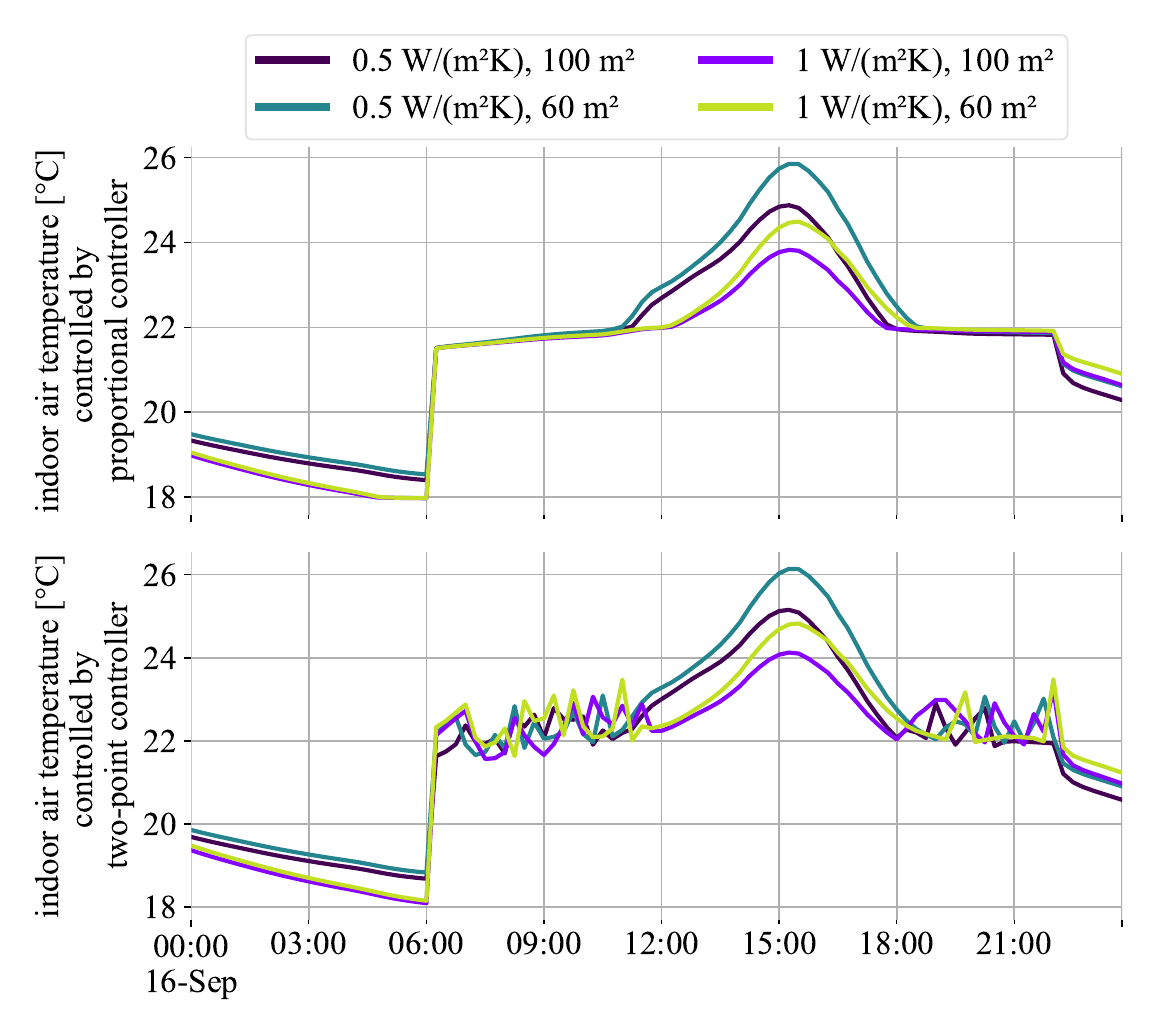}
    }
    \caption{(a) Daily average temperatures over one year for various simulations. (b) Indoor air temperatures over one day for different simulated buildings in Munich, using proportional and two-point controllers.}    
\label{fig:puzzles}
\end{figure}

\autoref{fig:mid_temp_data_diversity} presents the daily mean zone temperatures of five building variations over one simulation year. As no cooling is applied, the mean temperatures vary significantly depending on parameter combinations and solar radiation. \autoref{fig:controller_example_time_series} shows temperature curves for four variations on a September day, illustrating the effect of different controllers. The upper curves correspond to an internal P-controller, while the lower curves represent a two-point controller with night setback. Both controllers are set to 22~\textdegree C. While all buildings heat up similarly during the day, the two-point controller generates a distinct pattern by repeatedly switching the thermal source on and off to maintain the setpoint.

\begin{figure}[!htb]
    \centering
    \includegraphics[width=0.98\linewidth]{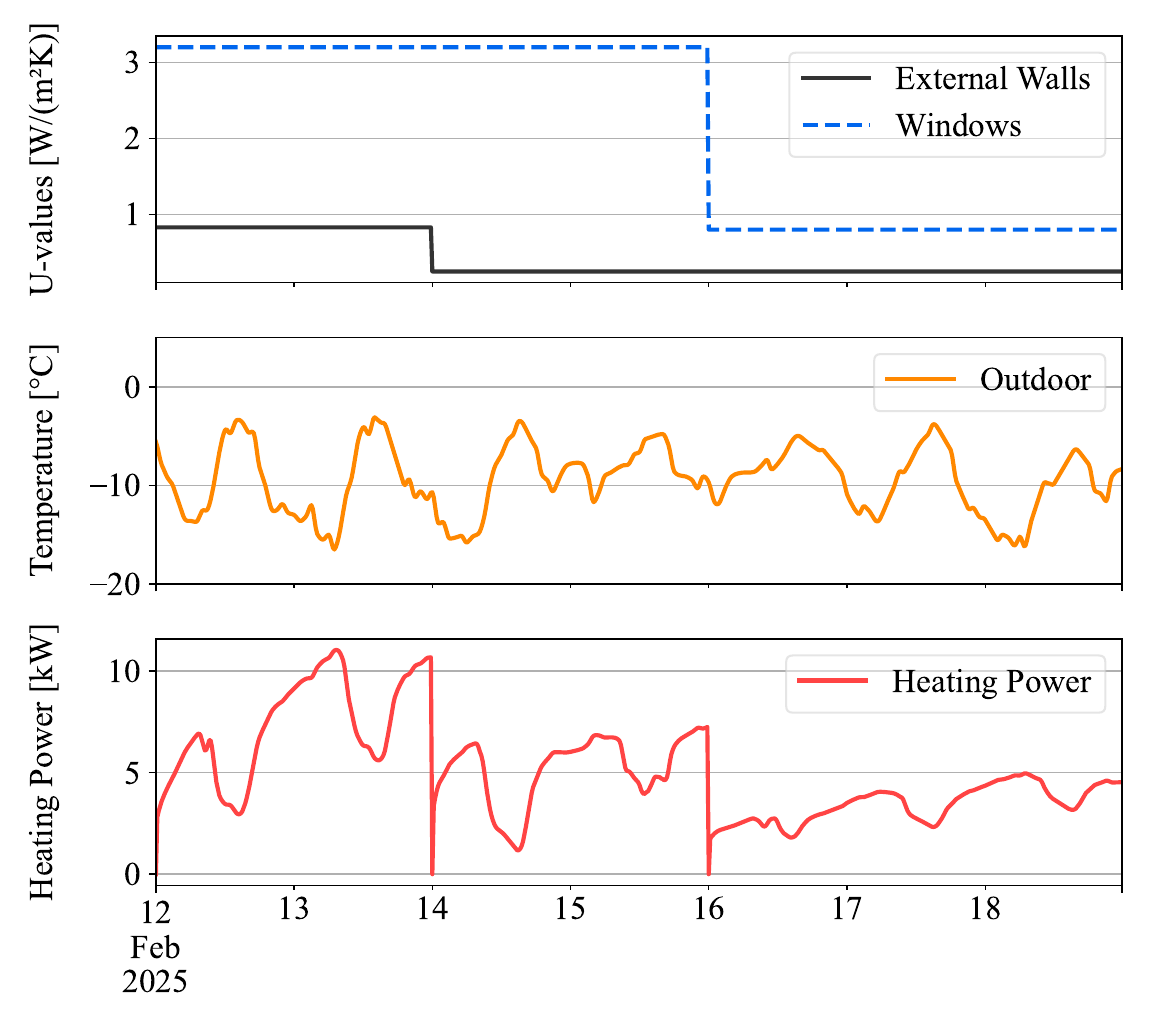}
    \caption{Demonstration of the impact of a retrofit on the heating energy demand by improving heat insulation of external walls and windows}
    \label{fig:model_parameter_change}
\end{figure}

The retrofitting capabilities of \texttt{BuilDa}, as described in Subsection \ref{cha:operational_changes} \nameref{cha:operational_changes}, are demonstrated in \autoref{fig:model_parameter_change}. Here, after a two days of simulation, the external wall is refurbished, reducing its U-value. Two days later, the windows are replaced with newer ones, which also decreases their U-value. The setpoint temperature for the zone is constantly at 20~\textdegree C. These two retrofitting measures lead to a significant reduction in the required heating power. In addition to external factors such as outside air temperature and solar radiative gains, they have a substantial impact on heating demand.

\begin{figure*}[!htb]
    \centering
    \includegraphics[width=0.98\linewidth]{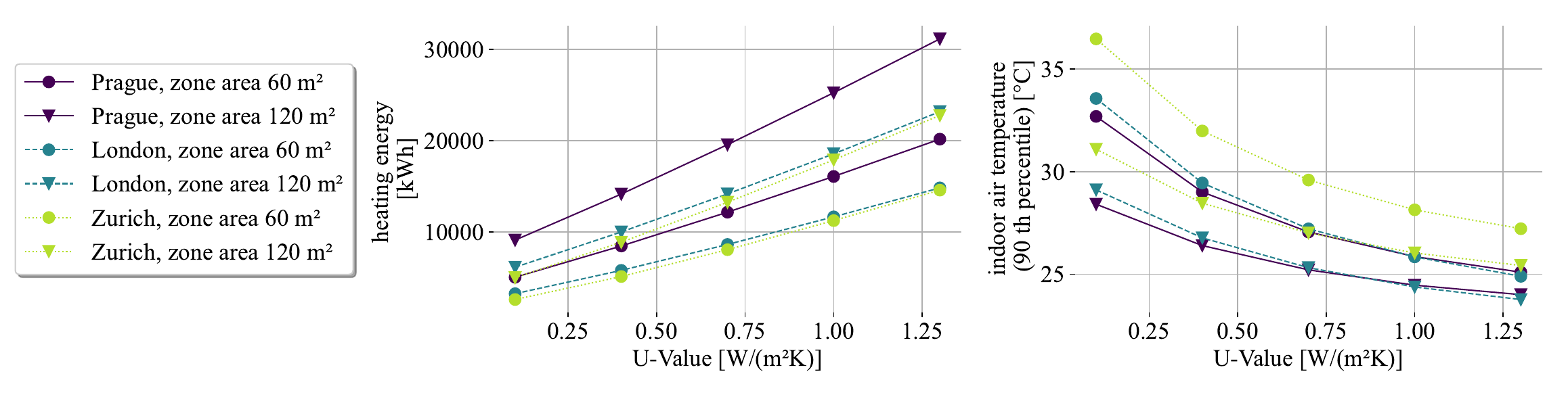}
    \caption{Relation between parameters U-value, weather, zone area and the output (left: heating energy, right: the 90th percentile of the indoor air temperature).}
    \label{fig:relation_parameter}
\end{figure*}

To further illustrate the impact of the building parameters, we show in \autoref{fig:relation_parameter} how parameters influence the energy consumption (left figure) and the 90th percentile of the indoor air temperature (right figure) of the buildings over one year. To compare different climates, we analyzed European locations with varying outdoor temperatures and solar radiation. Higher U-values increase heating energy demand while reducing indoor temperature peaks (90th percentile). Prague exhibits the highest heating energy consumption, followed by London and Zurich, while Zurich shows the highest temperature peaks. Larger zones (120 m²) require more heating energy but tend to have slightly lower temperature peaks than smaller zones (60 m²). These results highlight the inverse relationship between heating demand and peak indoor temperatures, better insulation (lower U-value) reduces heating needs but leads to higher peak temperatures. 

\subsection{Validation}
\label{cha:validation}
To ensure accuracy of the data, we validated the building model according to the ANSI/ASHRAE 140-2004 standard using the following test cases \cite{ASHRAE140}: TC600 and TC900 (fulfilled for annual heating and cooling demand),  
TC600FF and TC900FF (fulfilled for minimum, maximum, and mean temperature).  
The evaluation of the test cases is based on a comparison of the results with simulation outputs from a range of other established simulation tools. A result is considered valid if it falls within the corresponding reference range. The validation results and reference ranges are shown in \autoref{tab:validation_results}.

\begin{table}[htb]
    \footnotesize
    \centering
    \caption{Result of validation with ASHRAE test cases, heating/cooling demand in MWh, temperatures in °C}
    \label{tab:validation_results}
    \begin{tabular}{p{0.9cm} p{1.95cm} >{\centering\arraybackslash}p{1.0cm} p{0.7cm} p{0.7cm} >{\centering\arraybackslash}p{0.7cm}}
        \toprule
        \vspace{0.5em}Case & \vspace{0.5em}Metric & Results \texttt{BuilDa} & \vspace{0.5em}Min. & \vspace{0.5em}Max. & In Range \\
        \midrule
        TC600 & Heating Demand & 4.903 & 4.298 & 5.709 & \cellcolor{green!25}Yes \\ 
              & Cooling Demand & 6.59  & 6.137 & 7.964 & \cellcolor{green!25}Yes \\
        \midrule
        TC600FF & Average Temp. & 25.53 & 24.6 & 25.9 & \cellcolor{green!25}Yes \\
                & Minimum Temp. & -17.96 & -18.8 & -15.6 & \cellcolor{green!25}Yes \\
                & Maximum Temp. & 67.8  & 64.9 & 69.5 & \cellcolor{green!25}Yes \\
        \midrule
        TC900 & Heating Demand & 1.817 & 1.17 & 1.988 & \cellcolor{green!25}Yes \\
              & Cooling Demand & 2.99  & 2.132 & 3.415 & \cellcolor{green!25}Yes \\
        \midrule
        TC900FF & Average Temp. & 25.53 & 24.6 & 25.9 & \cellcolor{green!25}Yes \\
                & Minimum Temp. & -4.36 & -4.5 & -1.6 & \cellcolor{green!25}Yes \\
                & Maximum Temp. & 44.07 & 41.8 & 44.8 & \cellcolor{green!25}Yes \\
        \bottomrule
    \end{tabular}
\end{table}

\subsection{Transfer Learning Study} 
\label{cha:transfer_learning}

\subsubsection{Study Introduction}
\label{cha:Study_Introduction}

We now present an example of how to use \texttt{BuilDa} for TL research by pretraining various source models and fine-tuning them on a target. We consider the problem of source building selection, as it is crucial for successful TL in building thermal dynamics to choose the right source building for knowledge transfer \cite{pinto2022sharing, 2024Similarity}. Thorough studies on this topic are still rare, as research at the building-parameter level is challenging due to limited access to metadata (see Section \ref{sec:background} \nameref{sec:background}) or the need to manually prepare different physical simulation models \cite{CHAUDHARY2025115384}. We therefore demonstrate how changes on the building parameter level of source buildings can influence TL and how \texttt{BuilDa} can be utilized. This study is not aimed at comprehensively exploring all building parameters and their impact on TL for thermal dynamics. Instead, the study serves as a first step toward further, more in-depth parameter studies.

\subsubsection{Study Design}
\label{cha:Study_Design}

For the source buildings, we select five different building parameters that we want to vary with \texttt{BuilDa}. We chose parameters that have a strong influence on the building's thermal dynamics, namely, wall U-values, window U-values, the heat capacity of the walls, the area of the windows and the floor area. The selected parameter ranges correspond to the ranges of single-family houses depicted in \autoref{tab:rang_building_para}. \autoref{tab:tl_study_range} lists the resulting values. 

\begin{table}[htb]
    \centering
    \caption{Building parameter values of the simulation study}
    \label{tab:tl_study_range}
    \small 
    \begin{tabular}{p{2.3cm}ccc}
        \toprule
        \textbf{Parameter} & \textbf{Source Values} & \textbf{T\textsubscript{high}} & \textbf{T\textsubscript{low}}\\
        \midrule
        \parbox[t]{2.3cm}{Wall U-value\\ W/(m²K)} & 0.10, 0.75, 1.40 & 0.2 & 1.3\\ \midrule
        \parbox[t]{2.3cm}{Wall heat capacity\\ kJ/(m²K)} &  50, 250, 450  & 430 & 70 \\ \midrule
        \parbox[t]{2.3cm}{Window U-value\\ W/(m²K)} &  0.70, 1.90, 3.10  & 0.9 & 2.9 \\ \midrule
        \parbox[t]{2.3cm}{Floor area (m²)} &  60, 90, 120 & 110 & 70 \\ \midrule
        \parbox[t]{2.3cm}{Window-to-wall fraction} &  0.106, 0.128, 0.150  & 0.11 & 0.146 \\ 
        \bottomrule
    \end{tabular}
\end{table}

Note that the window area is defined by the window-to-wall fraction. Assuming buildings with a square floor plan, for a value of 0.106, this results in a minimal window area of 20 m² for the building with a floor area of 60 m², while a value of 0.15 results in a maximal window area of 40 m² for a building with 120 m² floor area, mirroring the TABULA values. Furthermore, the maximum value of the window U-value parameter was chosen to be 3.10, as this value better represents the characteristics of current building stock. We employ all possible parameter combinations, getting 243 different source buildings.

For the target buildings, we chose parameter combinations at the edge of the distribution, representing a building with high thermal inertia (T\textsubscript{high}) and low thermal inertia respectively (T\textsubscript{low}). The respective parameter values are listed in \autoref{tab:tl_study_range} and are distinct from the source building values.
By choosing targets on the edge of the distribution with opposing dynamics, effects on the building parameter level should become more easily apparent. We further assume a monolithic wall structure and the weather of Munich. We use the internal controller for heating with a lower and upper setpoint of 18\textdegree C (night) and 22\textdegree C (day), respectively. Except for the varied parameters, all other building parameters are identical between source and target buildings. The task we consider is the forecast of the indoor air temperature, depending on current conditions, i.e., the weather (direct and diffuse solar radiation, outside air temperature), the indoor air temperature, and the control signal of the thermal source. For the data-driven model, we chose an LSTM \cite{LSTM1997} with an additional fully connected layer to forecast the next 4 steps (1 hour), similar to the implementation in \cite{2025gentl}. We use the ADAM \cite{kingma2014adam} optimizer and the mean squared error as the cost function.   

We generate one year of data for the target buildings and for each of the 243 source buildings. We use the source building data to pretrain 243 distinct source models. For each source model, we first perform hyperparameter tuning for the hidden size, the number of layers, the learning rate and the batch size to get individual LSTM configurations, then we pretrain the models. Afterwards, each source model gets fine-tuned to the two targets, resulting in 486 fine-tuned models. The fine-tuned models inherit the hyperparameters from the pretrained source models. We rely on seasonal fine-tuning introduced by Raisch et al.\cite{2025gentl}, as each season (winter, spring, summer, autumn) influences the building dynamics differently, hence influences the fine-tuning success. For each season, we assume limited target data is available, e.g., 30 days in January for winter. The remaining data is used for evaluation, and best model selection is performed on this set. The RMSE of the selected model is reported as the test result. This design was deliberately chosen to avoid additional variance from dataset splits. The goal is to assess the relative impact of source building parameter variations on transfer learning performance rather than model generalization. The same procedure is applied to the months of April, July, and October, and the seasonal results are averaged to provide a robust overall performance estimate. 

To compare the prediction performance of the fine-tuned models, we also train target models from scratch. Training from scratch means the data-driven model is trained with the target data (30 days) without having a pretrained model as a starting point. We do this again for each of the 4 seasons and average the results. We also performed hyperparameter tuning for the two target models. We then compare the 486 fine-tuned models with the two target models trained from scratch.

\begin{figure*}[htb]
    \includegraphics[width=1 \linewidth]{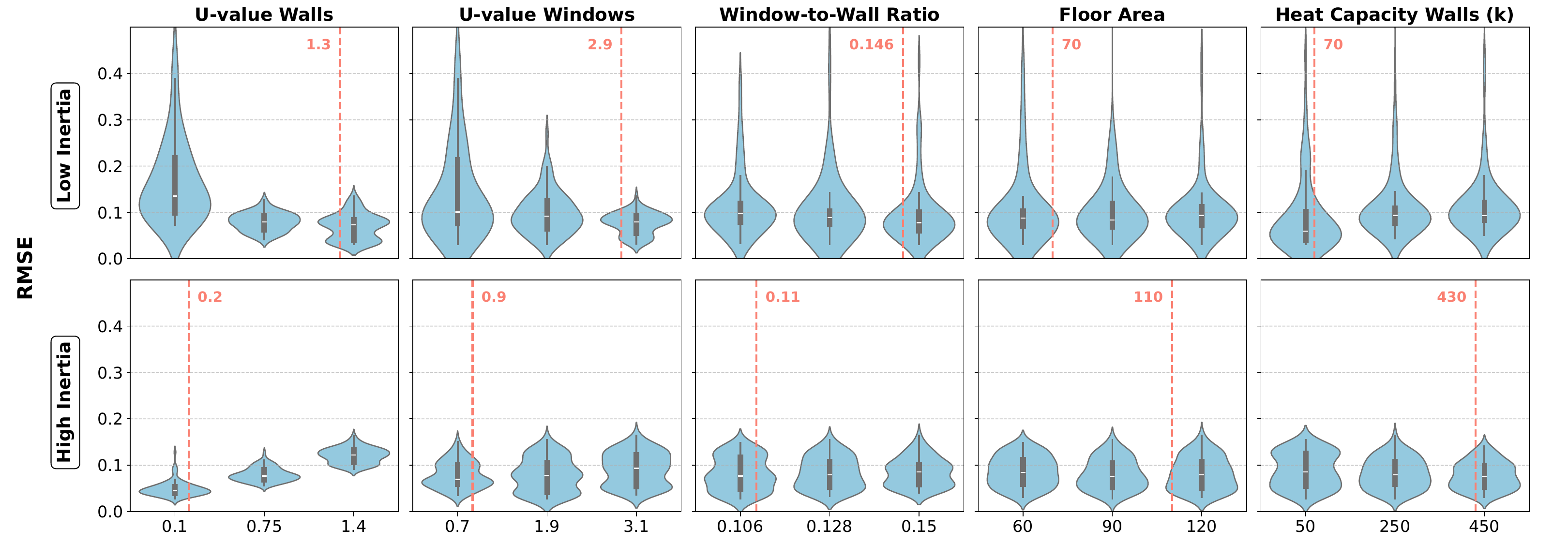}
    \centering
    \caption{The results of the fine-tuned models for both targets are shown (upper figure: low-inertia target, lower figure: high-inertia target). For each parameter value, the RMSEs of all fine-tuned models exhibiting this value (81 per value) are displayed as violin plots. Red lines indicate the actual target values of the target the models were fine-tuned to.}
    \label{fig:violin_plot}
\end{figure*}

\begin{figure*}[htb]
    \includegraphics[width=1 \linewidth]{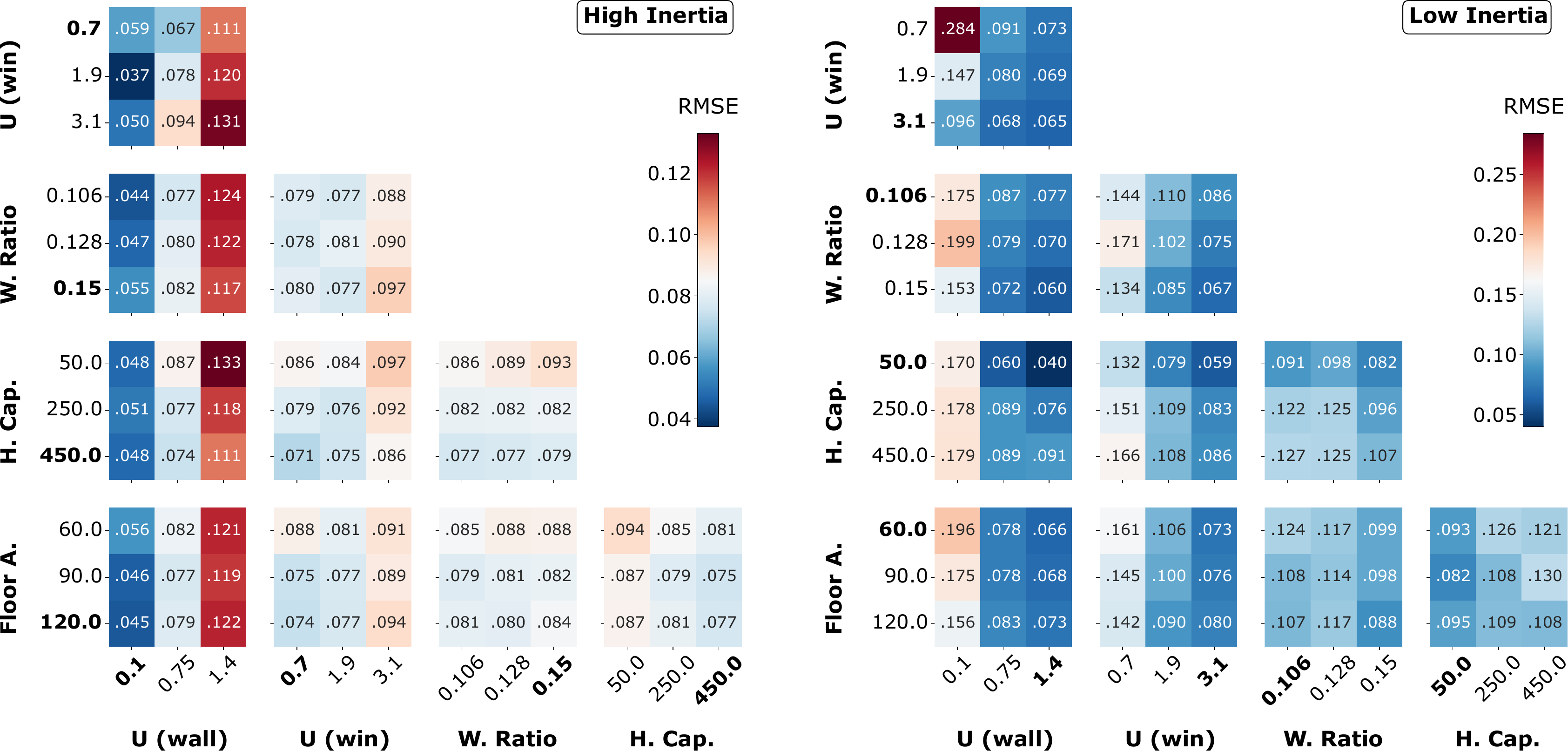}
    \centering
    \caption{Influence of parameter relations on the fine-tuning success of all models for both targets (left: high-inertia target; right: low-inertia target). Each square shows the averaged RMSE of all models exhibiting the respective two combined values (27 for each square). The values closest to the respective targets are marked bold.}
    \label{fig:heatmap}
\end{figure*}

\subsubsection{Study Results}
\label{cha:Study_Results}

\autoref{fig:violin_plot} shows the averaged RSME error after fine-tuning for both targets. To allow an easy comparison of the results and to avoid distortions, the upper tails of the low-inertia violin plot are truncated. The two worst-performing models (RMSEs 0.572 and 0.678) are included in the data but not fully shown. For the high-inertia target, a clear trend is only visible for the wall U-values. Wall U-values closer to the target improve fine-tuning performance. Other parameters show only very minor effects, and the overall spread of the mean RMSE is small, with values below 0.1 except for wall U-values. 

For the low-inertia target, clear trends are visible for wall U-values, window U-values, and wall heat capacity. Window size and floor area have minimal impact. The more pronounced trends seem to indicate, that for low-inertia targets, similar parameter values are more important for the fine-tuning success as for high-inertia targets. Further, the spread of the RMSEs is significantly greater than for the high-inertia target. This indicates that, for low-inertia targets, fine-tuning can still result in poor model performance, even when the parameter values of the source and target buildings are similar. This effect is particularly evident for wall heat capacity, window size, and floor area. It also suggests that fine-tuning to high-inertia targets is generally easier than to low-inertia targets

\autoref{fig:heatmap} Shows how parameter relationships affect TL performance for both targets. For the high-inertia target, parameter value combinations with wall U-values close to the target values dominate all other parameter combinations. For example, even if both floor area and wall heat capacity values match the target, they are less influential for TL success than a matching wall U-value. Conversely, mismatched wall U-values cause the highest overall prediction errors, although similar values for the other parameters can still slightly reduce the errors. 

For the low-inertia target, the dominant effect of the wall U-values remains visible, though the picture is less clear. When the wall U-value is combined with values close to the target for floor area, wall heat capacity, or window U-value, the fine-tuned models consistently perform very well. Interestingly, the window-to-wall fraction deviates from this trend, as larger window areas yield better performance. A window U-value close to the target also proves important, as shown by the best-performing combination of window U-value and wall heat capacity (average RMSE of 0.059). In comparison, the best-fitting combination of wall U-value and floor area achieves an averaged RMSE of 0.066, despite matching wall U-values. When neither wall nor window U-values match the target values, the worst-performing models occur (averaged RMSE of 0.284). Overall, this indicates that for low-inertia targets, wall and window U-values are the most important parameters for selecting suitable source buildings.

The RMSE of the models trained from scratch are \textbf{0.97} for T\textsubscript{high} and \textbf{1.08} for T\textsubscript{low}. As can be seen in \autoref{fig:violin_plot}, the TL models consistently achieve lower RMSEs, indicating superior performance of the TL approach.
This likely results from the similarities between source and target buildings, as only a fraction of all defining building parameters have been varied, and all buildings are exposed to the same weather. 

In summary, the results anecdotally demonstrate that TL models generally outperform models trained from scratch when only limited training data is available. Similar parameter values of the source to the target often result in a better model performance after fine-tuning. This effect seems to be especially important for low-inertia targets. Furthermore, fine-tuning appears easier for high-inertia targets, as indicated by consistently lower RMSEs and fewer outliers. However, it is still difficult to determine which source is best suited for a target, as unfavorable parameter value combinations can cause poor prediction performances, even when some parameter values closely match the target. Thus, similar building parameters do not necessarily imply similar thermal dynamics. 
Moreover, it remains unclear which building parameters are most critical for selecting an appropriate source building for successful TL. The results suggest that the wall U-value is particularly influential. For low-inertia targets, there is a implication, that next to the wall U-value, the window U-value is important to consider, especially in combination with the former parameter. This underlines that comprehensive and in-depth studies on building similarity are very much needed. We want to emphasize that, in addition to the selected parameters in our study, it is important to cover numerous other building parameter configurations, different climatic conditions, occupancy schedules, etc., as well as a larger set of target buildings to get a clear picture. \texttt{BuilDa} gives researchers a data generation tool to perform such comprehensive studies.

\section{Conclusion and Future Work}
\label{cha:conclusion}

We presented \texttt{BuilDa}, a highly flexible data generation framework for creating large amounts of high-fidelity thermal building dynamics time-series data, specifically designed for ML research. We described the variable physical building model and its features in detail, presented the overall architecture of the framework, and emphasized the distinct functionalities of \texttt{BuilDa}. The building model ensures physical consistency across all components, providing coherent and realistic thermal behavior. With \texttt{BuilDa} and its validated, high-fidelity building model, it becomes easy to simulate a wide range of building variations without requiring expert knowledge in building simulation. Building parameters, weather scenarios, operating conditions, and control strategies can be individually configured by the user, including the definition of retrofitting scenarios. Finally, we showcased examples of the heterogeneous data \texttt{BuilDa} can generate and demonstrated its use in a short yet robust TL study.

One possible research application is the pretraining of generalized Transformer architectures for building thermal dynamics forecasting, analogous to recent works on building load or power demand forecasting \cite{Hertel2023TansLoad, Gokhale2023TransformerDemand}. Such models can then be fine-tuned for individual buildings and serve as base models for control strategies like Model Predictive Control.
Further, \texttt{BuilDa} enables now comprehensive building similarity studies with a level of detail, which is not yet present in the TL community for building thermal dynamics. By allowing fine-grained parameter adjustments, systematic investigations of fine-tuning strategies, such as those explored by Chaudhary et al. \cite{CHAUDHARY2025115384}, become possible, but applicable to a broader range of buildings, including residential ones. This also supports deeper analyses of the relationships between building parameters and resulting thermal dynamics. The presented TL study in this paper lays the groundwork for future additional investigations that will explore these aspects in greater depth.

While we already made our framework available online (see Subsection \ref{cha:software_usage} \nameref{cha:software_usage}), we plan to provide key functionalities of \texttt{BuilDa} as an additional Python package, reducing barriers and allowing seamless integration of the variable building model into other research projects. By wrapping the FMU-interface with Gymnasium \cite{towers_gymnasium_2023}, this will also extend our framework towards RL research, similar to systems such as Sinergym \cite{Sinergym2021}. However, the variability of our physical model allows multiple buildings to be easily configured as RL training environments. 
In future work, we also plan to provide different base models representing various building types and multi-zone buildings, moving toward a universal and user-friendly ML research platform for building thermal dynamics.

\bibliographystyle{SageV}

\bibliography{buildabib}

\begin{acks}
Special thanks to Timo Germann and David Broos for their assistance with the project programming.
\end{acks}

\begin{funding}
This project is funded by the Helmholtz Association's Initiative and Networking Fund through Helmholtz AI. 
\end{funding}

\appendix

\section{Appendices} \label{app:quadratic}

\begin{figure*}[ht]
    \centering
    \includegraphics[width=0.78\linewidth]{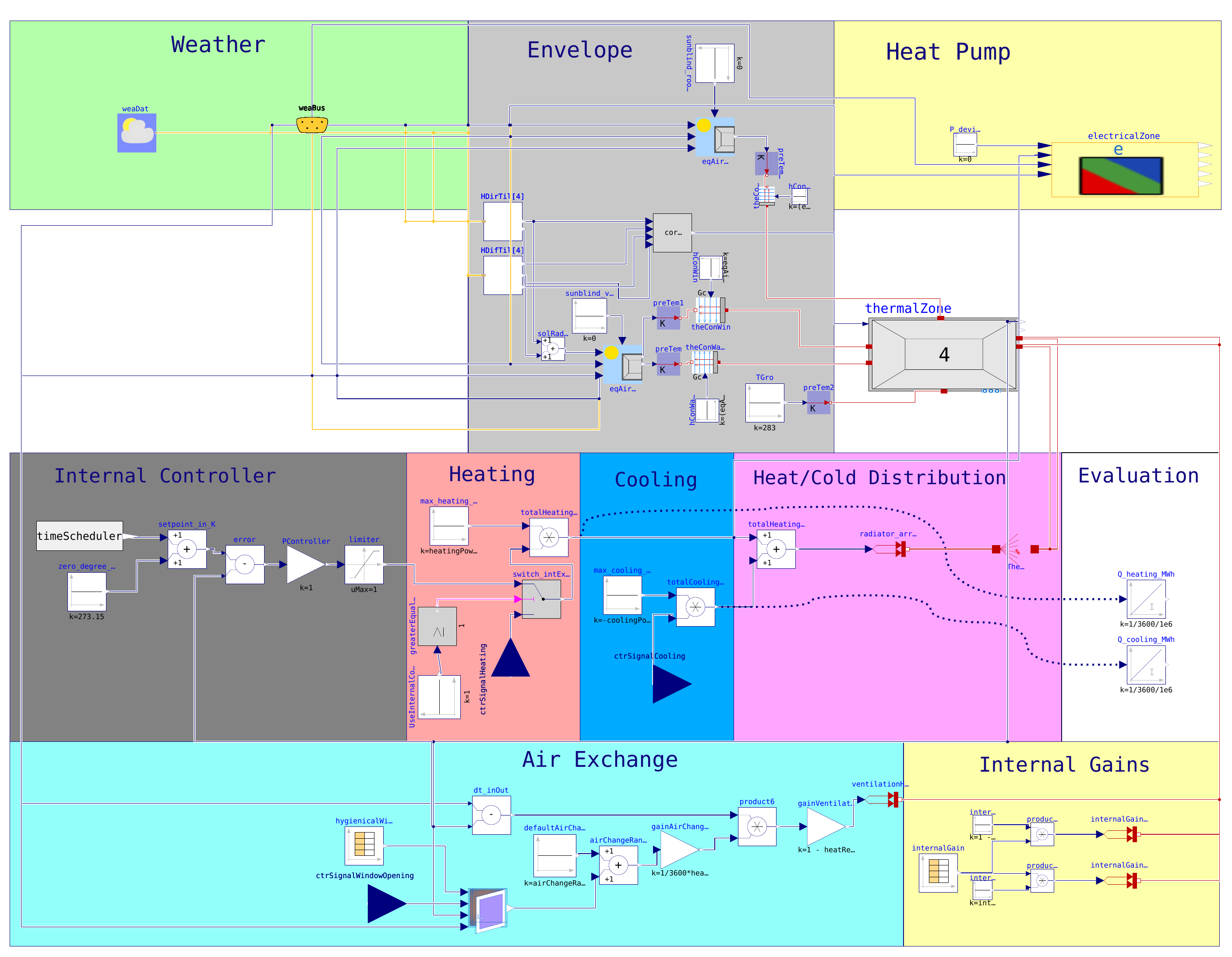}
    \caption{Overview of the implemented model in the Modelica editor with the following sections: 
    \textbf{Weather}, which provides an interface to the external weather file; 
    \textbf{Envelope}, the interface between the thermal zone and the environment; 
    \textbf{Heat Pump}, which calculates the required electric energy for heating using a heat pump; 
    \textbf{Internal Controller}, which controls the heat emission by the heating system based on the setpoint temperature; 
    \textbf{Heating}, which converts the controller output from normalized to absolute heating power and provides an external controller interface; 
    \textbf{Cooling}, which converts the controller output from normalized to absolute cooling power for the external cooling controller; 
    \textbf{Heat/Cold Distribution}, which divides the supplied heating/cooling energy into radiative and convective fractions and forwards it to the thermal zone; 
    \textbf{Evaluation}, which calculates the total heating and cooling demand during the simulation; 
    \textbf{Air Exchange}, which calculates heat losses induced by air exchange through infiltration, mechanical ventilation, or window opening, and provides an interface for window opening profiles or externally generated signals; 
    \textbf{Internal Gains}, which calculates heat gains induced by occupants or electrical devices, defined in an external file.
}

    \label{fig:model_overview}
\end{figure*}

\clearpage
\begin{table*}[htb]
\small
\caption{Variable building simulation parameters of \texttt{BuilDa}.}
\centering
\begin{tabular}{l|p{9.3cm}l}
Parameter name                   & Explication       & Unit                 \\  
\hline
        zone\_length, zone\_width                    & Length (from east to west) and width (from north to south) of zone                                  & m    \\
        n\_floors, floor\_height                    & Number and height of floor levels                                              & -, m    \\ 
        fAWin\_south    & Window-to-wall fraction on southern,                                    & -    \\ 
         \ \ \ \ \ \ \ \textit{(analogously for west, north, east)}                      &   \ \ \ \ \ \ \ western, norther and eastern wall & -    \\ 
        fATransToAWindow                 & Fraction of transparent window area to overall window area                      & -    \\ 
        fARoofToAFloor                   & Fraction of roof area to floor area (if inclined roof, this factor is $>1$)   & -    \\ 
        fAInt                            & Factor of exterior wall surfaces to interior wall surfaces (both sides)      & -    \\ 
        
        UExt                             & U-value of the external walls                                                & W/(m²K) \\
        \ \ \ \ \ \ \ \textit{(analogously for intWall, floor, roof)}         & \ \ \ \ \ \ \ internal wall, floor and roof                         & W/(m²K)    \\

        heatCapacity\_wall               & Heat capacity of the exterior walls related to its area                                                & J/(m²K) \\ 
        \ \ \ \ \ \ \ \textit{(analogously for intWall, floor, roof)}         & \ \ \ \ \ \ \ internal wall, floor and roof                         & J/(m²K)    \\
                heatCapacity\_furniture\_per\_m2 & Specific heat capacity of the furniture related to the floor area                                              & J/(m²K)                \\
        UWin                             & U-value of the windows                                                       & W/(m²K) \\ 
        thermalZone.gWin                & G-value of the windows            & - \\

        weaDat.fileName                  & Path(s) to weather file(s)                                                     & -    \\ 
        internalGain.fileName            & Path(s) to internal gain profile file(s)                                           & -    \\ 
        hygienicalWindowOpening.fileName& Path(s) to hygienical window opening profile file(s)                                & -    \\ 
        
        heatRecoveryRate                 & Heat recovery rate of ventilation system                                                                       & - \\
        airChangeRate                    & Air change rate due to ventilation system or infiltration                                                                          & - \\        
        roomTempLowerSetpoint            & Lower temperature setpoint (for night setback on internal controller)      & -    \\
        roomTempUpperSetpoint            & Upper temperature setpoint (for night setback on internal controller)      & -    \\
        UseInternalController             & Decides if model internal controller for heating is used                     & -    \\ 
        extWall\_C\_distribution   & Heat capacity distribution profile for external wall,    & -    \\
        \ \ \ \ \ \ \ \textit{(analogously for intWall, floor, roof)}         & \ \ \ \ \ \ \ internal wall, floor and roof                         & -    \\
        extWall\_R\_distribution       & Heat resistance distribution profile for external wall,        & -    \\
        \ \ \ \ \ \ \ \textit{(analogously for intWall, floor, roof)}         & \ \ \ \ \ \ \ internal wall, floor and roof                         & -    \\
        internalGainsConvectiveFraction& Fraction of internal gains that are convective                               & -    \\
        heatingConvectiveFraction         & Fraction of heating that is convective                                       & -    \\
        relative\_heatPump\_efficiency         & Efficiency of heat pump compared to reference heat pump
                                      & -    \\
        heatingCurve\_steepness         & Steepness of the heat curve; higher values indicate a greater increase in supply temperature relative to the difference between outside and inside temperatures. Typical range: 0.2 - 1.5                                       & -    \\
        \#extWall\_construction         & Parameter that can be used to configure one of the predefined component constructions (see Subsection \ref{cha:Building Parameters} \nameref{cha:Building Parameters})                                       & -    \\
\end{tabular}
\label{tab:allParameters}
\end{table*}

\end{document}